\documentclass[%
 superscriptaddress,
 amsmath,amssymb,
 aps,
 prb,
 twocolumn,
 longbibliography 
]{revtex4-2}

\usepackage{graphicx}
\usepackage{dcolumn}
\usepackage{bm}
\usepackage{xcolor}
\usepackage{tikz}
\usepackage{xfrac}
\usepackage{blindtext}
\usepackage{times}
\usepackage{multirow}
\usepackage{chemformula}

\usepackage{color}
\usepackage[colorlinks=true, urlcolor=blue, linkcolor=blue, citecolor=blue, pdftex]{hyperref}

\usepackage[normalem]{ulem}

\definecolor{darkblue}{HTML}{004D6B}
\definecolor{darkred}{HTML}{8c1515}
\definecolor{darkgreen}{HTML}{006400}
\usepackage{hyperref}
\hypersetup{
	colorlinks=true,
	urlcolor=darkred,
	citecolor=darkblue,
	linkcolor=darkred,
	breaklinks
}

\usepackage[caption=false]{subfig}

\graphicspath{{figures/}}

\newcommand{\be}{\begin{equation}}
\newcommand{\ee}{\end{equation}}

\newcommand{\bea}{\begin{eqnarray}}
\newcommand{\eea}{\end{eqnarray}}
\newcommand{\beal}{\begin{align}}
\newcommand{\eeal}{\end{align}}

\newcommand{\goto}{\rightarrow}

\renewcommand{\vec}[1]{\ensuremath{\mathbf{#1}}}

\newcommand{\abs}[1]{\left| #1 \right|} 
\newcommand{\avg}[1]{\left< #1 \right>} 

\begin{document}

\title{Frustrated Ferromagnetism of Honeycomb Cobaltates:\\
	Incommensurate Spirals, Quantum Disordered Phases, and Out-of-Plane Ising Order} 

\author{Yoshito Watanabe}
\affiliation{Department of Advanced Materials Science, University of Tokyo, Kashiwa 277-8561, Japan.}

\author{Simon Trebst}
\affiliation{Institute for Theoretical Physics, University of Cologne, 50937 Cologne, Germany.}

\author{Ciar\'an Hickey}
\affiliation{Institute for Theoretical Physics, University of Cologne, 50937 Cologne, Germany.}

\begin{abstract}
Recent experiments on the Co-based $d^7$ honeycomb materials BaCo$_2$(AsO$_4$)$_2$ and BaCo$_2$(PO$_4$)$_2$ have drawn renewed interest to XXZ models with competing interactions beyond the nearest neighbor exchange. While the case of dominant antiferromagnetic exchange has been extensively studied, the actual materials exhibit a ferromagnetic nearest neighbor exchange. Here we show that such a sign change of the dominant nearest neighbor coupling has substantial consequences on the phase diagram. In particular, the nature of the quantum disordered phase of interest  
changes from a plaquette valence bond crystal to a long-range entangled spin liquid phase. By means of complementary numerical simulations, based on exact diagonalization and the pseudo-fermion functional renormalization group, we demonstrate the formation of a gapless spin liquid state at the heart of the ferromagnetic phase diagram in the isotropic Heisenberg limit, which gives way to out-of-plane Ising magnetic order upon inclusion of XXZ anisotropy.
The magnetically ordered phases surrounding this quantum disordered region are much less sensitive to the sign change of the dominant interaction and can be understood from their classical analogs.  
We briefly comment on the relevance of our results for Co-based $d^7$ honeycomb materials.
\end{abstract}
\maketitle


\noindent

Magnetic frustration, the inability of a system to simultaneously satisfy all of its competing interactions, can carry dramatic consequences for the low-temperature properties of magnetic materials \cite{Moessner2001}. Indeed, in some cases, it can actually prevent long-range magnetic order even at zero temperature \cite{Balents2010}. The resulting non-magnetic states can come in many different flavors, such as valence bond crystals (VBCs), symmetry-breaking phases with a rigid arrangement of singlet bonds, or quantum spin liquids (QSLs), highly entangled phases with fractionalized excitations and emergent gauge structures \cite{Savary16}. The possibility of observing such quantum states in materials has inspired decades of searches for ideal candidate materials and relevant microscopic models. 

Recently, Co-based $d^7$ honeycomb materials have experienced a renewed wave of interest \cite{Nair2018, Zhong2020, 2205.15262,2209.15510}. This was largely driven by theoretical suggestions \cite{Liu2018,Sano2018} that such materials may host significant bond-dependent Ising interactions, making them potential candidate materials for realizing the spin liquid physics of the Kitaev honeycomb model \cite{Kitaev06,Hermanns18,Trebst22}. 
However, at least for the cobaltate materials \ch{BaCo2(AsO4)2} and \ch{BaCo2(PO4)2}, it seems increasingly likely that their original theoretical description, in terms of an XXZ $J_1$-$J_2$-$J_3$ model, better captures their essential physics \cite{2205.15262,2211.03737,Das2021,Maksimov2022}. 
Notably, the materials are believed to host a ferromagnetic (FM) nearest-neighbor $J_1$ interaction \cite{Regnault1977,Rastelli1979,Das2021,Maksimov2022,2205.15262,2211.03737}, 
while theoretical studies have almost exclusively focused on the antiferromagnetic (AFM) $J_1$ case
\cite{Fouet2001,Albu2011,Reuther2011,Mosadeq2011,Li2012,Mezzacapo2012,Kalz2012,Gong2013,Zhang2013,Zhu2013,Ganesh2013,Li2014,Maksimov2016}. In the classical limit \cite{Rastelli1979,Mulder2010,Li2012b,Shimokawa2019}, the rich phase diagrams of the two models are related by a simple mapping, meaning they share the same underlying classical physics \cite{Fouet2001}.
 However, this mapping breaks down in the quantum spin-$1/2$ limit, opening the possibility for unconventional quantum magnetism in a frustrated ferromagnet, distinct from its antiferromagnetic counterpart.

This has led us to study the phase diagram of the honeycomb $J_1$-$J_2$-$J_3$ XXZ model with a fixed FM $J_1$ interaction, from the isotropic Heisenberg limit to the fully anisotropic XY limit. 
At first glance, the majority of the parameter space for the quantum model contains the same magnetically ordered phases that appear in the classical limit, including relatively simple phases such as FM and N\'eel ordered phases, as well as more complex incommensurate spiral ordered phases (see Fig.~\ref{fig:Classical_Sketch} below). However, along the classical phase boundary between the FM phase and nearby incommensurate phases, there is a high degree of frustration which inhibits conventional magnetic order. Indeed, for the quantum model we find a narrow region which, for isotropic Heisenberg interactions, does not exhibit long-range magnetic order, but instead contains a dense spectrum of low-lying excitations with a gap scaling highly suggestive of a {\em gapless} ground state -- a highly entangled state that should be distinguished from the gapped (and moderately entangled) plaquette valence bond crystal (PVBC) put forward in the context of AFM $J_1$ couplings \cite{Mosadeq2011,Albu2011,Zhu2013,Ganesh2013}. Upon introducing an XXZ anisotropy the gapless spin liquid phase for FM coupling rapidly disappears and, as one approaches the XY limit, a phase with long-range out-of-plane $zz$-Ising order \cite{Li2014, Zhu2014} develops, consistent with previous studies in the XY limit \cite{Bishop2014}, for which the FM and AFM models become equivalent again. Our insights are based on a combination of both exact diagonalization (ED) and pseudo-fermion functional renormalization (pf-FRG) calculations which allow us to provide a general overview of the XXZ model and, in particular, focus on the properties of the candidate QSL phase that appears near the Heisenberg limit.


\textit{Model and Symmetries.--} 
The starting point of our study is the $J_1$-$J_2$-$J_3$ XXZ model on the honeycomb lattice with a fixed FM $J_1<0$ and the same XXZ anisotropy $\Delta\leq 1$ for all bonds. The Hamiltonian can be written compactly as,
\begin{equation}
	H = \sum_{n=1}^3 J_n \sum_{\avg{i,j}_n} \left( S_i^x S_j^x + S_i^y S_j^y + \Delta S_i^z S_j^z \right)  \, ,
	\label{eq:Hamiltonian}
\end{equation} 
where $\Delta=1$ corresponds to the $J_1$-$J_2$-$J_3$ Heisenberg model with full SU(2) symmetry, and $\Delta=0$ corresponds to the $J_1$-$J_2$-$J_3$ XY model with a reduced U(1) symmetry.

\begin{figure}[t]
\includegraphics[width=\columnwidth]{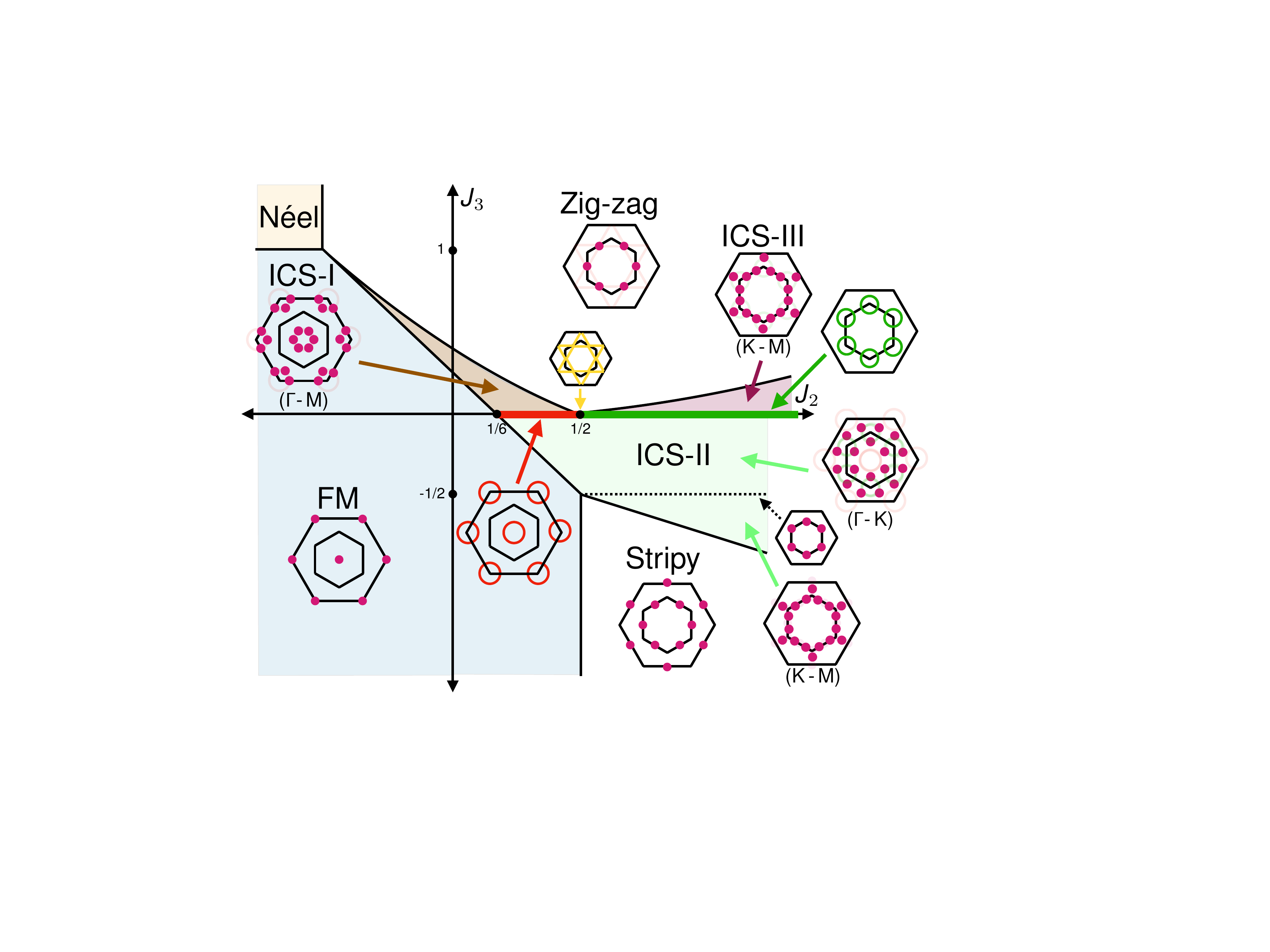}
\caption{
	\textbf{Classical phase diagram} of the $J_1$-$J_2$-$J_3$ Heisenberg model with {\it ferromagnetic} $J_1<0$.
	Besides the ferromagnetically (FM) and zig-zag (ZZ) ordered phases there are three different spin-spiral ordered phases
	with incommensurate wave vectors (ICS-I/II/III), which are here defined via the locations of their ordering wavevectors in momentum space (insets).
	On the $J_3=0$ line there is a spin-spiral liquid regime for $J_2/|J_1| \geq 1/6$, with classically degenerate rings of possible ordering wavevectors.
}
\label{fig:Classical_Sketch}
\end{figure}

In the classical limit, it's possible to do a spin inversion operation on the spins of just one sublattice, say the $A$ sublattice, such that $\vec{S}_{i\in A} \goto -\vec{S}_{i\in A}$ and $\vec{S}_{i\in B} \goto \vec{S}_{i\in B}$. This changes the parameters of the Hamiltonian such that $J_1 \goto - J_1$, $J_2 \goto J_2$ and $J_3 \goto - J_3$. Thus a model with FM $J_1$ gets mapped to a model with AFM $J_1$, accompanied by a change in the sign of $J_3$. As a result, the full classical phase diagrams of the two models share the same topography \cite{Fouet2001}. However, in the quantum model, for any finite $\Delta  >0$, this mapping breaks down as it is not possible to change the sign of the spins of just one sublattice (a full spin inversion, $\vec{S}_i \goto -\vec{S}_i$, can only be achieved globally with a time-reversal symmetry operation). In the pure XY limit, i.e.~$\Delta=0$, since $S_i^z$ drops out of the Hamiltonian a simple global spin rotation, a $\pi$-rotation about the $z$-axis, results in $\left(S_i^x,S_i^y\right) \goto \left(-S_i^x,-S_i^y\right)$. Thus, in this limit, the quantum phase diagrams of the two models, with FM and AFM $J_1$, are again related to one another.


\textit{Classical Phase Diagram.--} 
To set the stage, let us recap the classical phase diagram for the model which was mostly worked out some four decades ago \cite{Rastelli1979}, identifying a number of ordered phases that are either collinear or coplanar. A summary is given in Fig.~\ref{fig:Classical_Sketch} where, besides adopting more descriptive names for the various phases, we show representative spin structure factors. This phase diagram is identical for both the Heisenberg and XY limits (as well as everything in between), with the only difference being that the spins lie within the $xy$-plane for $0\leq \Delta <1$. 

The main features can be understood by starting from the physics of just the $J_1$-$J_2$ model (i.e. the horizontal line at the center of Fig.~\ref{fig:Classical_Sketch}):
(i) For $J_2/\abs{J_1} < 1/6$, the ground state is a simple ferromagnet, with ordering wavevector $\vec{Q}^\star = \Gamma$ (and $\Gamma^\prime$). Adding a small $J_3$ does not destabilize the ground state. Indeed, this state dominates a large portion of the phase diagram, including, unsurprisingly, the entire quadrant in which all couplings are FM. 
(ii) For $1/6 < J_2/\abs{J_1} < 1/2$, there is an entire degenerate ring of possible ordering wavevectors centered about the $\Gamma$ (and $\Gamma^\prime$) point, a manifestation of a residual, subextensive ground-state manifold at zero temperature. This leads to classical ``spiral spin liquid" physics \cite{Bergman2007,Shimokawa2019} at finite temperature, in which the system fluctuates between states with different $\vec{Q}^\star$. An infinitesimal AFM $J_3>0$ (FM $J_3<0$) explicitly breaks the ring degeneracy and selects a finite set of incommensurate ordering wavevectors that lie on the $\Gamma$-$M$ ($\Gamma$-$K$) high-symmetry lines, with the resulting incommensurably ordered phases labelled as ICS-I (II). 
(iii) Exactly at $J_2/\abs{J_1}=1/2$, the rings connect and there is a full line degeneracy in momentum space. An infinitesimal AFM $J_3>0$ selects the $M$ points, resulting in zig-zag magnetic order. On the other hand, a FM $J_3<0$ again selects incommensurate ordering wavevectors along the $\Gamma$-$K$ high-symmetry line. 
(iv) For $J_2/\abs{J_1} > 1/2$, the line degeneracy splits and there are again degenerate rings of possible ordering wavevectors, this time centered about the $K$ and $K^\prime$ points. An infinitesimal AFM $J_3>0$ (FM $J_3<0$) selects incommensurate ordering wavevectors that lie on the $K$-$M$ ($\Gamma$-$K$) high-symmetry line, with the resulting phases labelled as ICS-III (II). Note that, contrary to Ref.~\cite{Rastelli1979}, we have distinguished ICS-II and ICS-III as two distinct phases as the energy exhibits a discontinuity as one crosses from one to the other.

\begin{figure*} 
\includegraphics[width=0.9\linewidth]{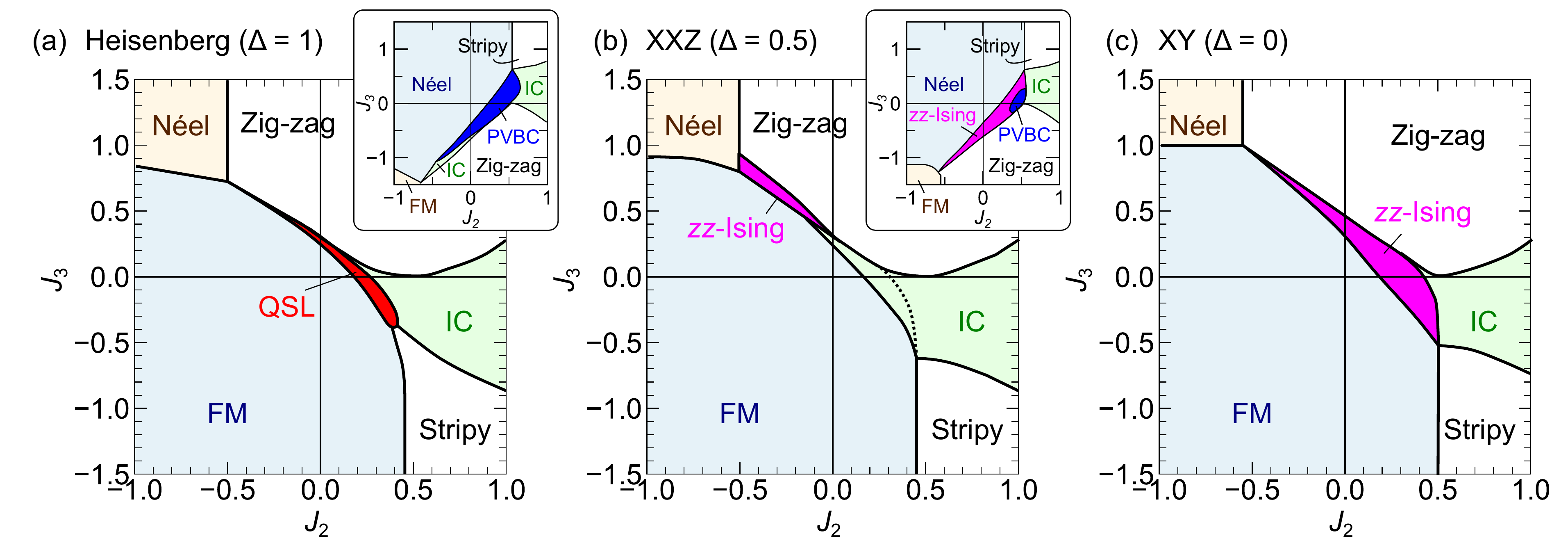}
\caption{
{\bf Quantum phase diagrams for varying spin anisotropy $\bf\Delta$} of model \eqref{eq:Hamiltonian} with ferromagnetic nearest-neighbor exchange $J_1$. 
The different panels show, from left to right, (a) the isotropic Heisenberg model, (b) an anisotropic XXZ model, and (c) the XY model.
As a comparison, the insets in (a) and (b) show the distinct phase diagrams for antiferromagnetic $J_1$.
The determination of phase boundaries and phase assignments are done using the critical scale $\Lambda_c$ and structure factor
obtained by pf-FRG calculations (except for the assignment of the PVBC in the inset, which is labelled according to previous works \cite{Mosadeq2011,Albu2011,Zhu2013,Ganesh2013}). 
}
\label{fig:PD_summary}
\end{figure*}
\begin{figure}[b] 
\includegraphics[width=\columnwidth]{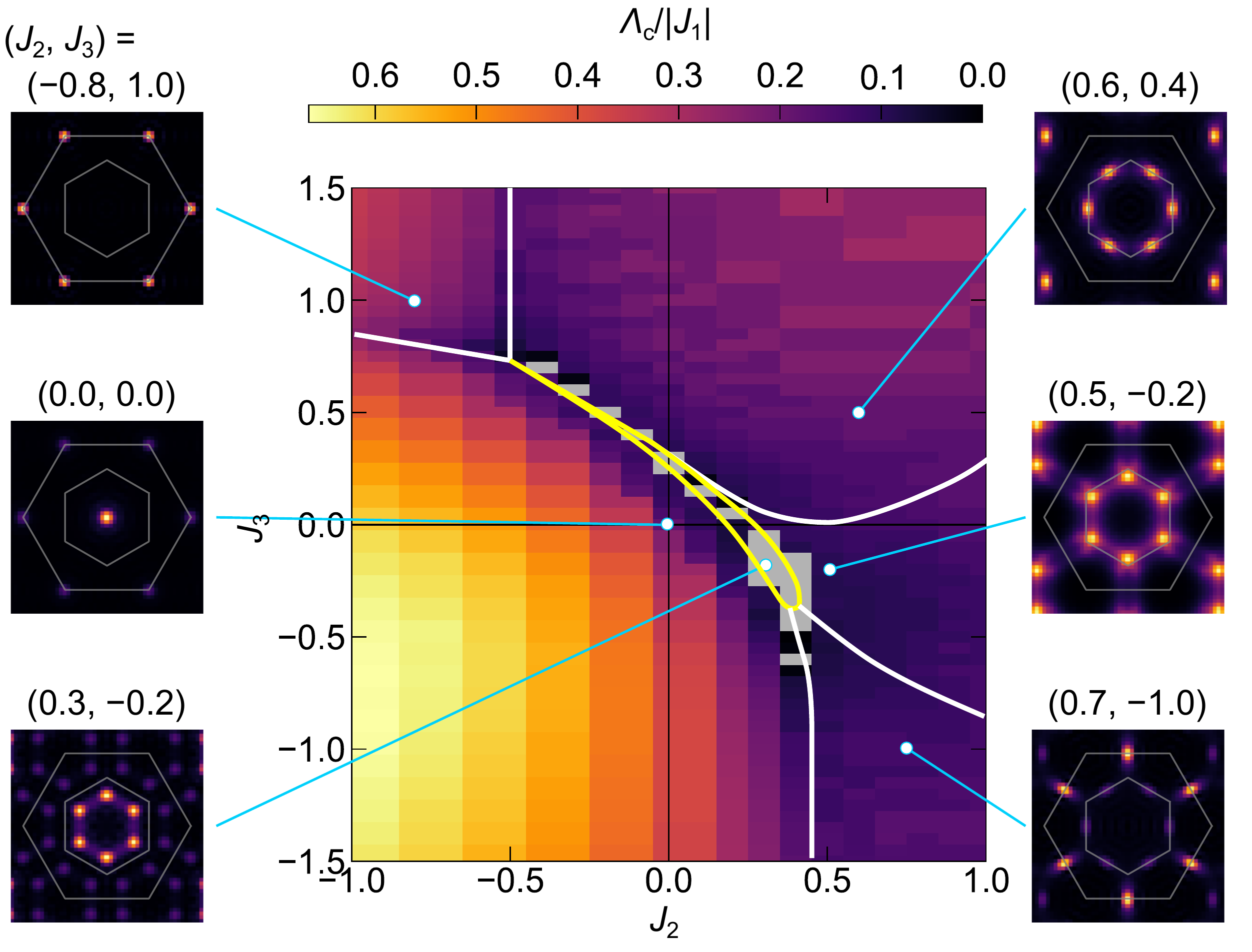}
\caption{
{\bf Finite-temperature phase diagram} of model \eqref{eq:Hamiltonian} in the Heisenberg limit. 
Color coded is the break-down scale $\Lambda_c$ of the flow parameter in pf-FRG calculations,
indicating the onset of magnetic ordering.  
A finite $\Lambda_c$ can be roughly translated into an ordering temperature scale via $T_c \propto \Lambda_c \pi/2$ \cite{Iqbal2016,Buessen2016}. 
The grey boxes indicate parameters for which no flow breakdown could be identified down to the lowest scales 
(typically $\Lambda/|J_1| \sim 0.01$), pointing to the formation of a quantum disordered state.
The solid line indicates the zero-temperature phase diagram of Fig.~\ref{fig:PD_summary}(a).
The insets show static spin structure factors as a comparison to the classical phase diagram of Fig.~\ref{fig:Classical_Sketch}. 
}
\label{fig:FM_Heisenberg}
\end{figure}

\textit{Quantum Phase Diagrams.--} 
Let us now turn to the quantum physics of model \eqref{eq:Hamiltonian}, which we have broadly mapped out in the phase diagrams of Fig.~\ref{fig:PD_summary} using pf-FRG calculations \cite{Reuther2010}, a technique which is particularly well suited to capture incommensurate spin spiral order in quantum magnets \cite{Baez2017} but can also identify quantum disordered regimes \cite{Buessen2018}. 
Our simulations are based on the \texttt{PFFRGSolver.jl}~\cite{kiese2021,PFFRGSolver} software package.
Shown in Fig.~\ref{fig:PD_summary} are phase diagrams for three values of spin anisotropy $\Delta$, including the Heisenberg and XY limits, as well as one value in between ($\Delta=0.5$)
\footnote{
It's important to note that, in the pure Heisenberg limit, aspects of the model have been studied before, with, for example, exact diagonalization (ED) being used to study the properties of the ordered phases \cite{Fouet2001}, as well as the phase diagram along the line $J_2=J_3$ \cite{Li2012}. On the other hand, in the pure XY limit, there have been numerous studies \cite{Bishop2014,Li2014, Zhu2014} investigating the structure of the phase diagram and trying to pin down the nature of the phases that appear therein (most of these studies focus on the AFM $J_1$ case but, as discussed earlier, the AFM and FM cases can be mapped one to the other).
}.
 A notable departure from the classical phase diagram is that the region of highest frustration shifts from the line of $J_3=0$ 
 towards the boundary between the FM and adjacent incommensurate ordered phases. In the Heisenberg limit, this region lacks long-range magnetic order for the quantum model, signified by a lack of a spin-symmetry breaking flow breakdown within pf-FRG, see Fig.~\ref{fig:FM_Heisenberg} and the Supplemental Material (Fig.~\ref{fig:flow_examples}). On the other hand, in the XY limit, this region possesses a long-range magnetic order that has no classical counterpart in the classical phase diagram. This non-classical quantum phase has a N\'eel-type Ising order with dominant $\avg{S_i^zS_j^z}$ spin-spin correlations.

\begin{figure} 
\includegraphics[width=\columnwidth]{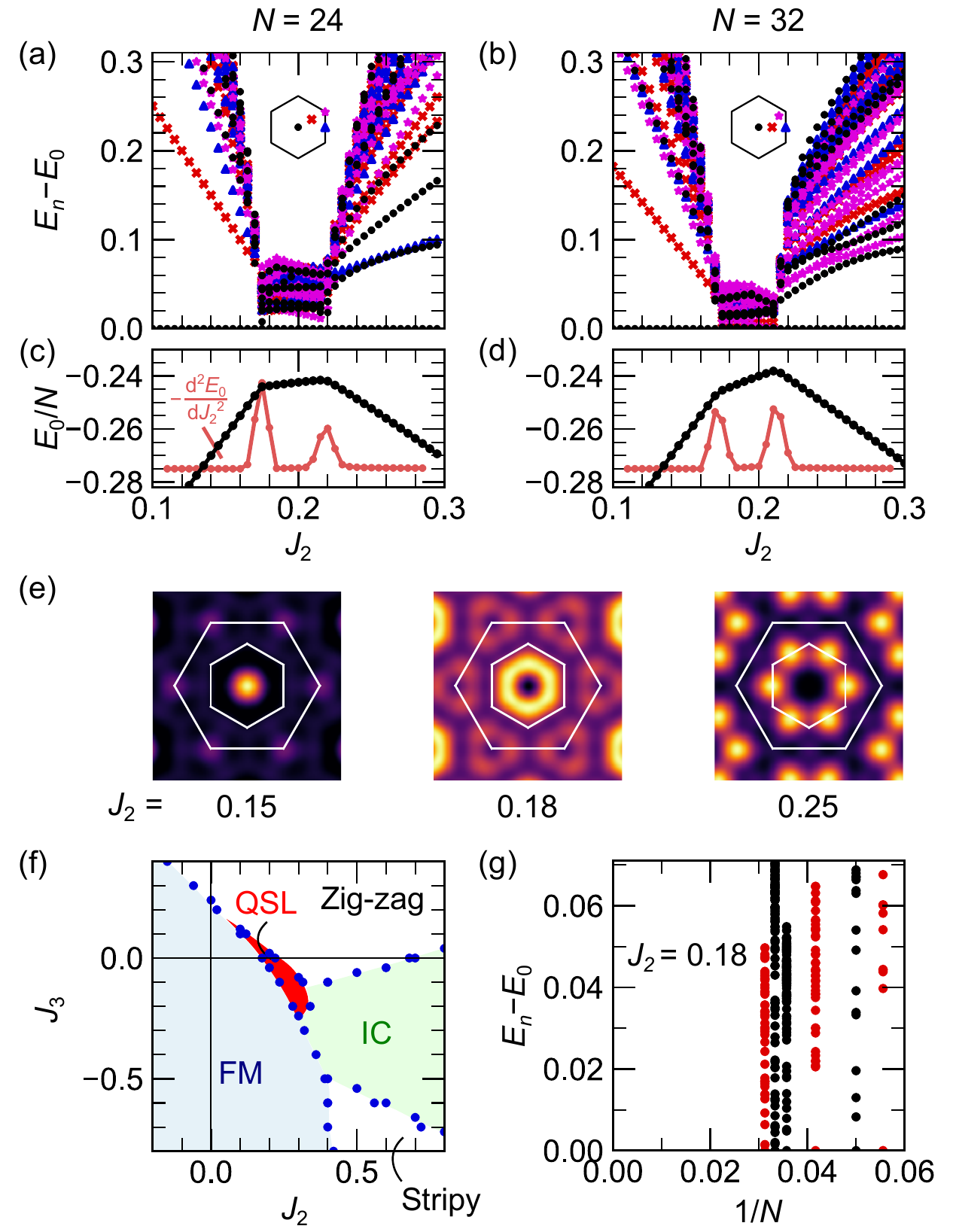}
\caption{
{\bf Spectral signatures of spin liquid physics.}
(a, b) Low-energy spectrum of $J_1$-$J_2$ Heisenberg model for (a) $N = 24$ and (b) $N = 32$ sites.
The distinct momentum sectors used for the ED calculation are shown in the first BZ in each inset.
The six lowest states from each momentum and spin inversion sector are plotted.
(c, d) The ground state energy $E_0$ and its second derivative.
(e) Static spin structure factor $S(\bm{q})$ and (f) phase diagram from ED calculations for $N=24$ sites.
(g) System size dependence of low-energy excited states. The red dots indicate the C$_3$ symmetric clusters with $N=18,24,32$ sites, the black dots indicate clusters of size $N=20,28,30$.
}
\label{fig:ED_results}
\end{figure}

\textit{Quantum spin liquid in Heisenberg limit.--} In order to gain more insight into the nature of the non-magnetic phase in the Heisenberg limit, we turn to ED calculations on various clusters of up to $N=32$ spins, with the phase diagram shown in Fig.~\ref{fig:ED_results}(f) for the fully symmetric $N=24$ site cluster. The structure of the phase diagram qualitatively agrees with that obtained by pf-FRG, with again a narrow phase hugging the classical phase boundary between the FM and incommensurate orders. For simplicity we focus on the line $J_3=0$, i.e.~we restrict ourselves to the $J_1$-$J_2$ Heisenberg model (see Supplemental Material section \ref{sec:Stability} for the effects of a finite $J_3$). 

The spin structure factors, shown in Fig.~\ref{fig:ED_results}(e), are consistent with a non-magnetic intermediate state, with no sharp peaks visible for $J_2=0.18$ but rather the weight is evenly distributed in a ring about the $\Gamma$ point, reminiscent of the underlying classical spiral spin liquid physics (though unlike the classical model, this feature is observed even with a finite FM $J_3$, see Fig.~\ref{fig:J2_cut}). This is further supported by the lack of a clear tower of states in the energy spectrum (see Fig.~\ref{fig:tos}), which, when present, signifies the spontaneous breaking of spin rotational symmetry. The simplest possibility for this non-magnetic phase is that, though it does not break spin rotational symmetry, it instead spontaneously breaks lattice symmetries. Indeed, in the analogous regime of the model with AFM $J_1$, a non-magnetic phase also appears. The nature of the phase there has been hotly debated but, as mentioned earlier, the most prominent candidate is a PVBC phase, a non-magnetic phase that spontaneously breaks translational symmetry. However, such a phase is an unlikely candidate for the model with FM $J_1$ as ferromagnetic nearest-neighbor interactions suppress the formation of local singlets. Indeed, a direct calculation of the dimer-dimer correlation function confirms that dimer correlations are extremely weak, an order of magnitude lower than that found in the PVBC phase with AFM $J_1$ (see Fig.~\ref{fig:supp3}). 

The most intriguing clue as to the nature of the phase is revealed by inspecting the low-energy spectrum. In Fig.~\ref{fig:ED_results}(a) and (b), we see that there is a high density of states at the lowest energies, with states across all momenta contributing. Even more dramatically, in Fig.~\ref{fig:ED_results}(g), the energy gap decreases, and the density of states at low energies increases, with increasing system size. This suggests that the ground state is gapless in the thermodynamic limit \footnote{Note that Ref.~\cite{Fouet2001} studied the same model, $J_1$-$J_2$ Heisenberg model on the honeycomb lattice with FM $J_1$ and AFM $J_2$, using exact diagonalization. There, they focused on the parameter point $J_2/|J_1|=0.25$ and conjectured that the ground state is a gapped quantum spin liquid. However, this parameter point lies outside of our candidate gapless quantum spin liquid region and instead lies within the ordered region of the phase diagram. This can be most clearly seen in Fig.~\ref{fig:J2_cut} where we show a cut at $J_2/|J_1|=0.25$ with varying $J_3$.}. Such a scenario would further rule out the possibility of a valence bond crystal, or a fully trivial paramagnet (which is an alternative possibility allowed by the Lieb-Schultz-Mattis-Hastings-Oshikawa theorem \footnote{The Lieb-Shultz-Mattis-Hastings-Oshikawa theorem states that for a model with spin rotational and translational symmetries, and an odd number of spin-$1/2$ per unit cell, the ground state must either be gapless, spontaneously break symmetry, or possess topological order. However, in our model, on the honeycomb lattice, we have an even number of spin-$1/2$ per unit cell, which means it's possible to have a ground state that is gapped, symmetric and topologically trivial (and thus has a unique ground state).}), as these states are gapped. Rather, the combination of an absence of spin rotational symmetry breaking and a gapless spectrum points toward a QSL ground state. 

\begin{figure} 
\includegraphics[width=\columnwidth]{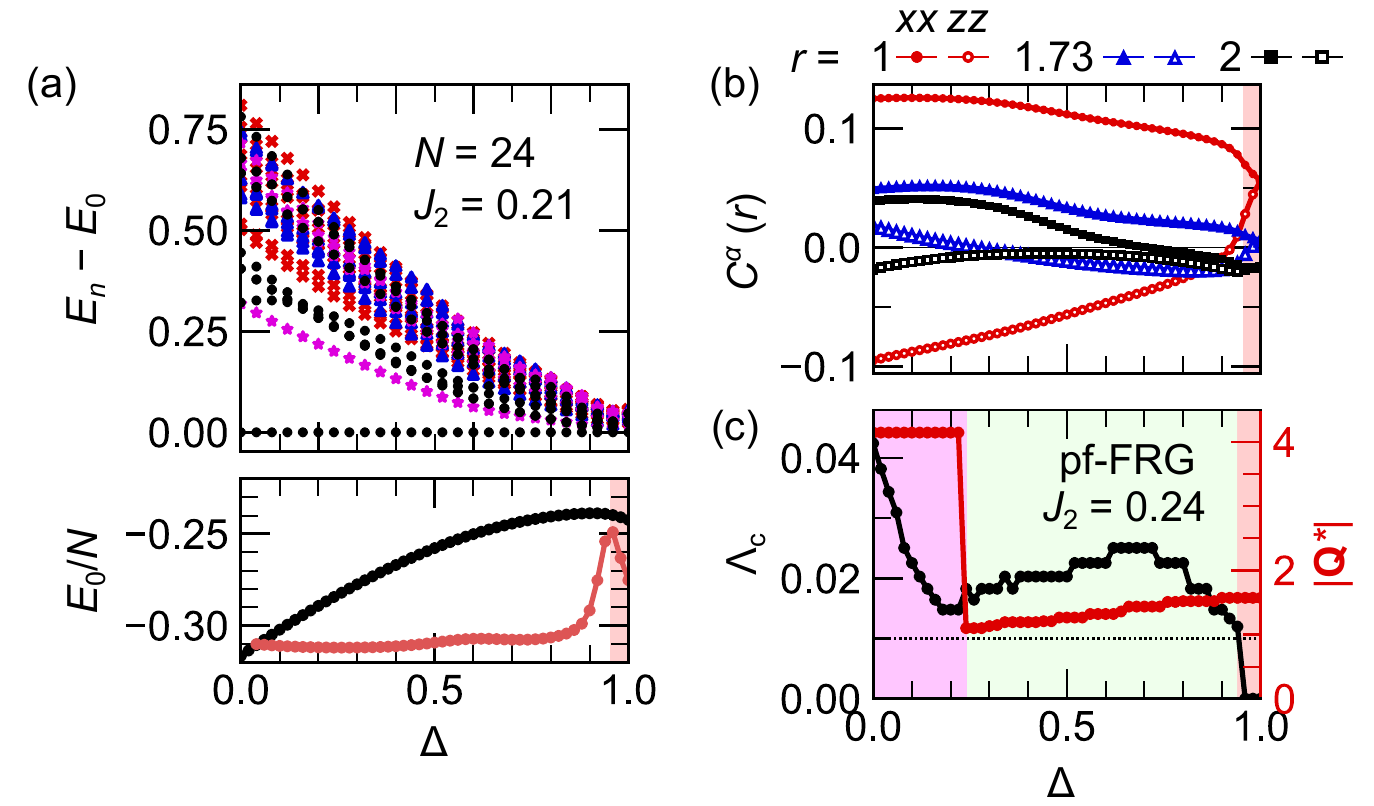}
\caption{
{\bf Formation of zz-Ising order and XY limit.}
(a) Low-energy spectrum of FM $J_1$--$J_2$ XXZ model.
(b) Real-space spin-spin correlations $C^{\alpha}(r) = \left<S^\alpha(0)S^\alpha(r)\right>$.
(c) Critical scale $\Lambda_c$ and the absolute value of the dominant wave-vector in the structure factor $|\vec{Q}^\star|$ obtained by the pf-FRG calculation.}
\label{fig:delta_scan}
\end{figure}

\textit{From Heisenberg to XY.--} Breaking SU(2) symmetry by turning on a finite XXZ anisotropy, $\Delta<1$, quickly destabilizes the candidate QSL. Fig.~\ref{fig:delta_scan} illustrates this for a representative parameter point. In ED, the dense low-energy spectrum rapidly lifts and there is a peak in the second derivative of the ground state energy, indicating a phase transition, at a value of $\Delta=0.96$. Moving further to the XY limit, Fig.~\ref{fig:delta_scan}(b) shows how the real-space spin-spin correlations evolve, with the nearest-neighbor $zz$ correlations quickly turning negative, i.e.~AFM, and eventually reaching a value close in magnitude to the positive (FM) $xx$ and $yy$ correlations. This is rather remarkable given that in the XY limit the nearest-neighbor interactions are purely FM and the Hamiltonian does not involve the $z$-component of the spin. 

The pf-FRG results of Fig.~\ref{fig:delta_scan}(c) reveal the full picture, with again the non-magnetic region (signified by the absence of a flow breakdown, i.e.~breakdown scale $\Lambda_c=0$) restricted to a narrow window about the Heisenberg limit, with critical $\Delta_c= 0.96$. Below that, there are two magnetically ordered phases. For $0.24 < \Delta < 0.96$, the spins order within the $xy$-plane at an incommensurate wavevector, in agreement with the classical expectation. On the other hand, for $0 < \Delta < 0.24$, the $zz$ component of the magnetic susceptibility dominates and peaks at the $\Gamma^\prime$ high symmetry points (see Fig.~\ref{fig:supp4}). This indicates an out-of-plane $zz$ Ising ordered state, with a N\'eel ordered pattern in real space. 

On a technical level, the ability of pf-FRG to capture the unusual physics of the out-of-plane Ising ordered phase in the pure XY limit is a further testament to its ability to accurately describe magnetically long-range ordered phases, even, in this case, one without a classical counterpart. It also adds further evidence supporting the absence of quantum disordered phases, either VBCs or QSLs, in the honeycomb lattice extended XY model.


\textit{Dicussion.--} The most interesting region of the phase diagram uncovered by this work lies within a narrow window hugging the classical phase boundary between FM and incommensurate orders, as well as a narrow window about the Heisenberg limit where we find a quantum disordered phase. With regard to the Co-based $d^7$ materials \ch{BaCo2(AsO4)2} (BCAO) and \ch{BaCo2(PO4)2} (BCPO), an experimental study on powder samples of BCPO estimates $J_1$-$J_2$-$J_3$ values very close to this quantum disordered phase \cite{Nair2018}. However the estimate for the XXZ anisotropy moves it firmly away. Similarly, studies of BCAO suggest both a negligible $J_2$ and significant XXZ anisotropy, again moving it away from the disordered region and placing it within the ordered regions. Outside of the cobaltates, the layered honeycomb magnet \ch{FeCl_3} can be described by a simple FM $J_1$-AFM $J_2$ Heisenberg model, supplemented by weak interlayer interactions. However, there, the high spin, $S=5/2$, of the \ch{Fe} moments pushes the physics towards the classical regime and spin spiral liquid physics living there \cite{Gao2022}.    

On the theoretical front, there are several possible candidates for the potentially gapless QSL degrees of freedom. A classification of projective symmetry groups (PSGs) for QSLs with a $\mathbb{Z}_2$ gauge structure has revealed a number of possibilities with protected Dirac crossings of the fermionic partons \cite{Lu2011}. For QSLs with a $U(1)$ gauge structure, gapless fermions, either in the form of Dirac cones, or a full Fermi surface, are necessarily required to generate a stable phase. However, on a bipartite lattice such as the honeycomb lattice, the Dirac $U(1)$ spin liquid is more susceptible to monopole proliferation as a trivial, symmetric monopole operator always exists \cite{Song2020}. 

Intriguingly, a recent study of classical XY spiral spin liquids with approximately circular rings of degenerate ordering wavevectors showed that the physics there can be understood using the language of $U(1)$ tensor gauge theory \cite{Yan2021}. As shown in Figs.~\ref{fig:Classical_Sketch} and \ref{fig:PD_summary}, this situation occurs for the classical $J_1$-$J_2$ model precisely near the boundary between the FM and incommensurate order where our candidate QSL appears in the Heisenberg limit of the quantum spin-$1/2$ model. It would be interesting to explore, in future work, whether the quantum model exhibits any such tensor gauge theory physics, in analogy to the classical model.


\acknowledgements
We thank T. Lorenz and J. Reuther for discussions, and L. Gresista for his technical support in the pf-FRG calculation
using the \texttt{PFFRGSolver.jl}~\cite{kiese2021,PFFRGSolver} software package. 
Y.W. is supported by JSPS through the Program for Leading Graduate Schools (MERIT).
We acknowledge partial funding from the DFG within Project-ID 277146847, SFB 1238 (projects C02, C03) and
JSPS KAKENHI (No. JP19H0588), Japan.
The numerical simulations were performed on the JUWELS cluster at the Forschungszentrum Juelich
and the Noctua2 cluster at PC$^2$ in Paderborn. 



%



\section*{Supplementary Material}

To supplement our discussion in the main text of the nature of the putative gapless QSL phase
in proximity to the boundary between the FM and incommensurate spin spiral phases,
we here provide six supplemental plots with additional numerical data which are referred to in the main text.

\makeatletter
\renewcommand \thesection{S\@arabic\c@section}
\renewcommand\thetable{S\@arabic\c@table}
\renewcommand \thefigure{S\@arabic\c@figure}
\makeatother
\setcounter{figure}{0}

\section{Representative pf-FRG flows}

\begin{figure}[h!]
\includegraphics[width=\columnwidth]{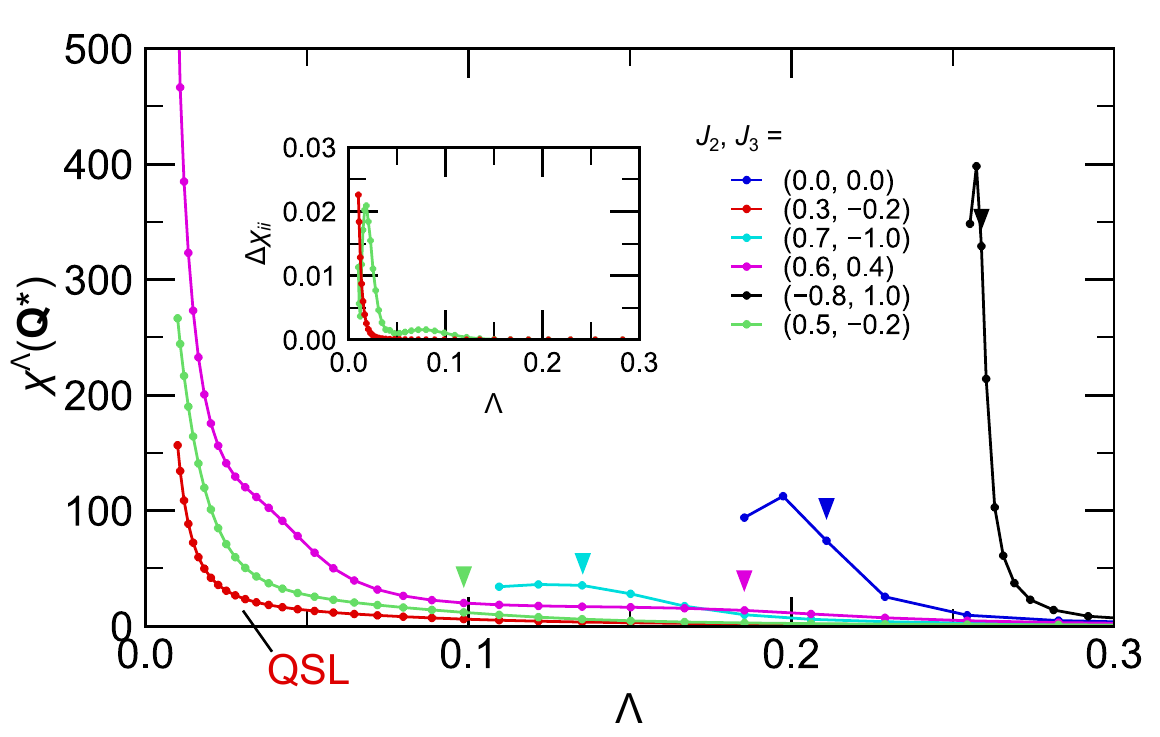}
\caption{
{\bf Structure factor flows from pf-FRG} for various coupling parameters/phases.
Flow breakdowns, marked by triangles, indicate the formation of magnetic ordering.
The inset shows differences in on-site spin-spin correlation flow
between different lattice sizes, i.e.,
$\Delta\chi_{ii}\equiv |\chi^{L=15}_{ii} - \chi_{ii}^{L=11}/\chi_{ii}^{L=11}|$.
}
\label{fig:flow_examples}
\end{figure}

As a technical supplement to our pf-FRG calculations used to determine the phase diagram in Fig.~\ref{fig:PD_summary} 
of the main manuscript and to identify suppressed magnetic ordering scales in Fig.~\ref{fig:FM_Heisenberg}, we present 
typical FRG flows of the magnetic structure factor in Fig.~\ref{fig:flow_examples}.
Magnetic ordering is indicated in these flows by a breakdown of a smooth flow/curve.
Such a breakdown is easily identifiable if a sharp peak forms in the flow; 
in such cases, we identify the critical scale $\Lambda_c$ (plotted in Fig.~\ref{fig:FM_Heisenberg} of the main text) 
as the last data point before that peak.
In more subtle cases, where only a weak shoulder-like feature emerges in the flow, 
we define the local minimum in the second derivative of the flow as the breakdown point.
The signature of the flow breakdown in the IC phase [$(J_2, J_3) = (0.5, -0.2)$ in the figure] is subtle in the structure factor flow
but becomes clearer in the scaling behavior of on-site correlation $\Delta\chi_{ii}$,
as shown in the inset of Fig~\ref{fig:flow_examples}.

\begin{figure}[h!]
\includegraphics[width=\columnwidth]{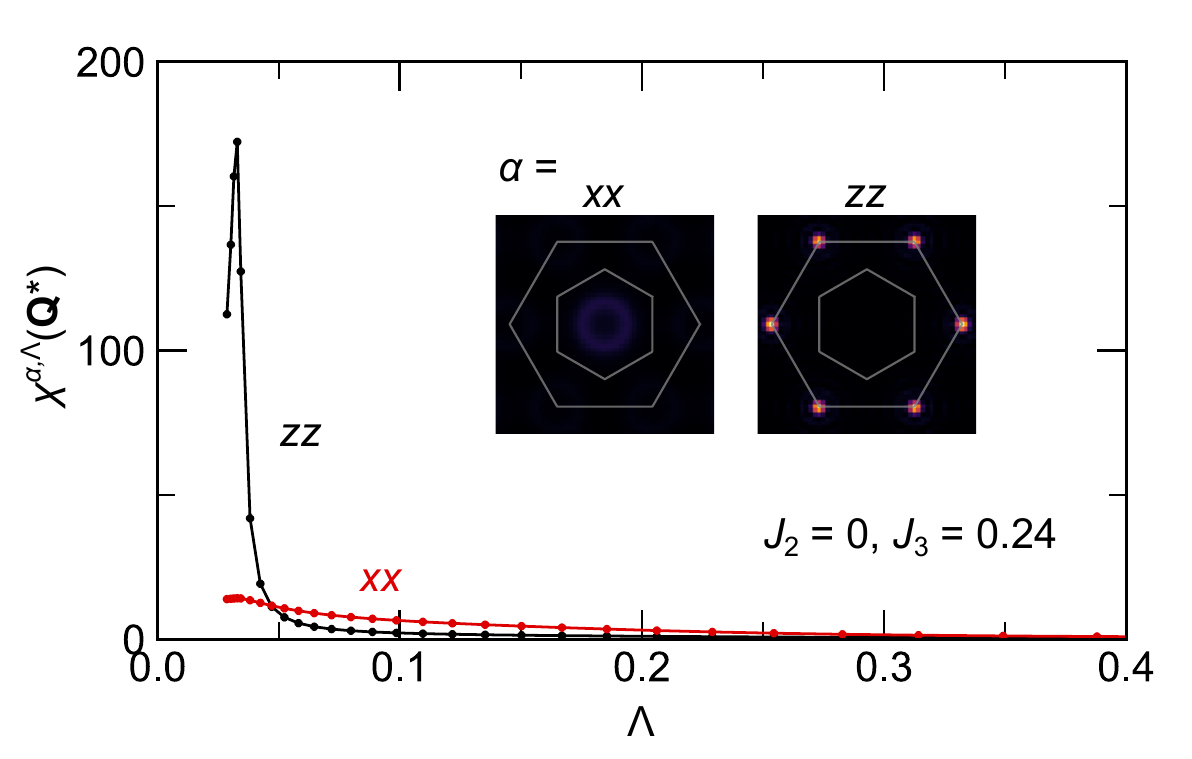}
\caption{
{\bf pf-FRG flows in the Ising ordered phase.}
The main panel shows the structure factor flow for its $xx$ and $zz$ components,
with the dominant wave vectors $\vec{Q}^\star$ chosen independently for the two components.
The inset shows the structure factor at $\Lambda_c$ for the two components (from which $\vec{Q}^\star$ is chosen).
}
\label{fig:supp4}
\end{figure}

Of particular technical interest to pf-FRG practitioners might be the fact that our pf-FRG simulations
for the anisotropic model indeed allows to see the formation of $zz$-ordering in a system that only 
exhibits XY couplings. This is shown in Fig.~\ref{fig:supp4} where the magnetic $zz$ structure factor 
shows a clearly discernible breakdown in this XY model.

\section{Stability of quantum spin liquid phase} \label{sec:Stability}

\begin{figure}[!t]
\includegraphics[width=\columnwidth]{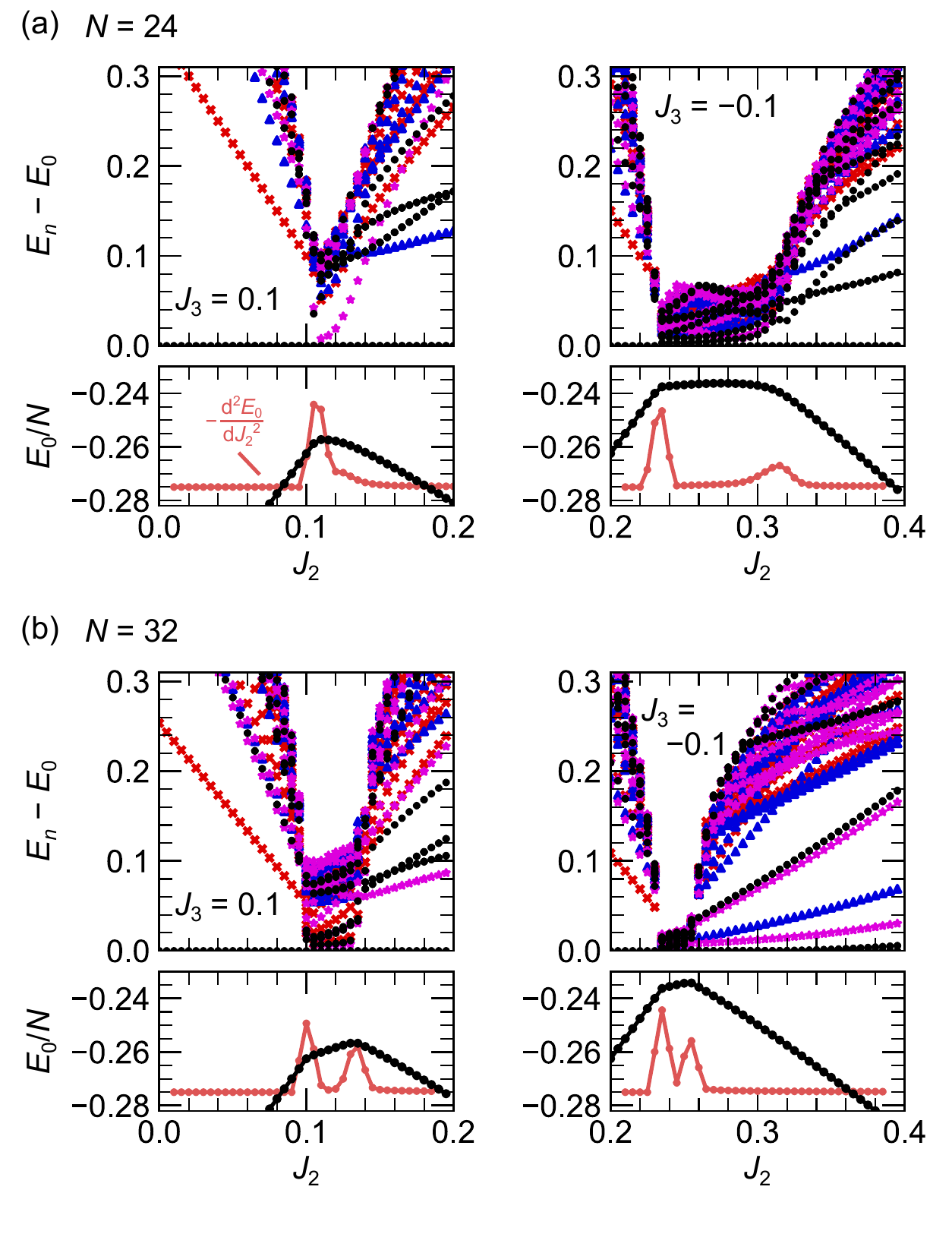}
\caption{
{\bf Stability of the QSL phase for finite $J_3$}.
Shown are the energy-spectrum and ground state energy, calculated by ED for (a) $N = 24$ and (b) $N = 32$ sites,
along a horizontal cut of varying $J_2$.
}
\label{fig:supp1}
\end{figure}

To demonstrate the stability of the spin liquid forming along the boundary of the ferromagnetic phase, 
we here present a number of additional line cuts through the phase diagram of Fig.~\ref{fig:PD_summary}
in the main text. Figure \ref{fig:supp1} shows a horizontal cut (varying $J_2$) for antiferromagnetic and
ferromagnetic $J_3$ in the left/right panels of Fig.~\ref{fig:supp1}, respectively. 
Similar to the $J_3=0$ scan of Fig.~\ref{fig:ED_results} in the main text an intermediate phase with a collapsing energy
spectrum is clearly discernible, flanked by clear signatures in the second derivative of the ground-state energy.

\begin{figure}[t]
\includegraphics[width=.9\columnwidth]{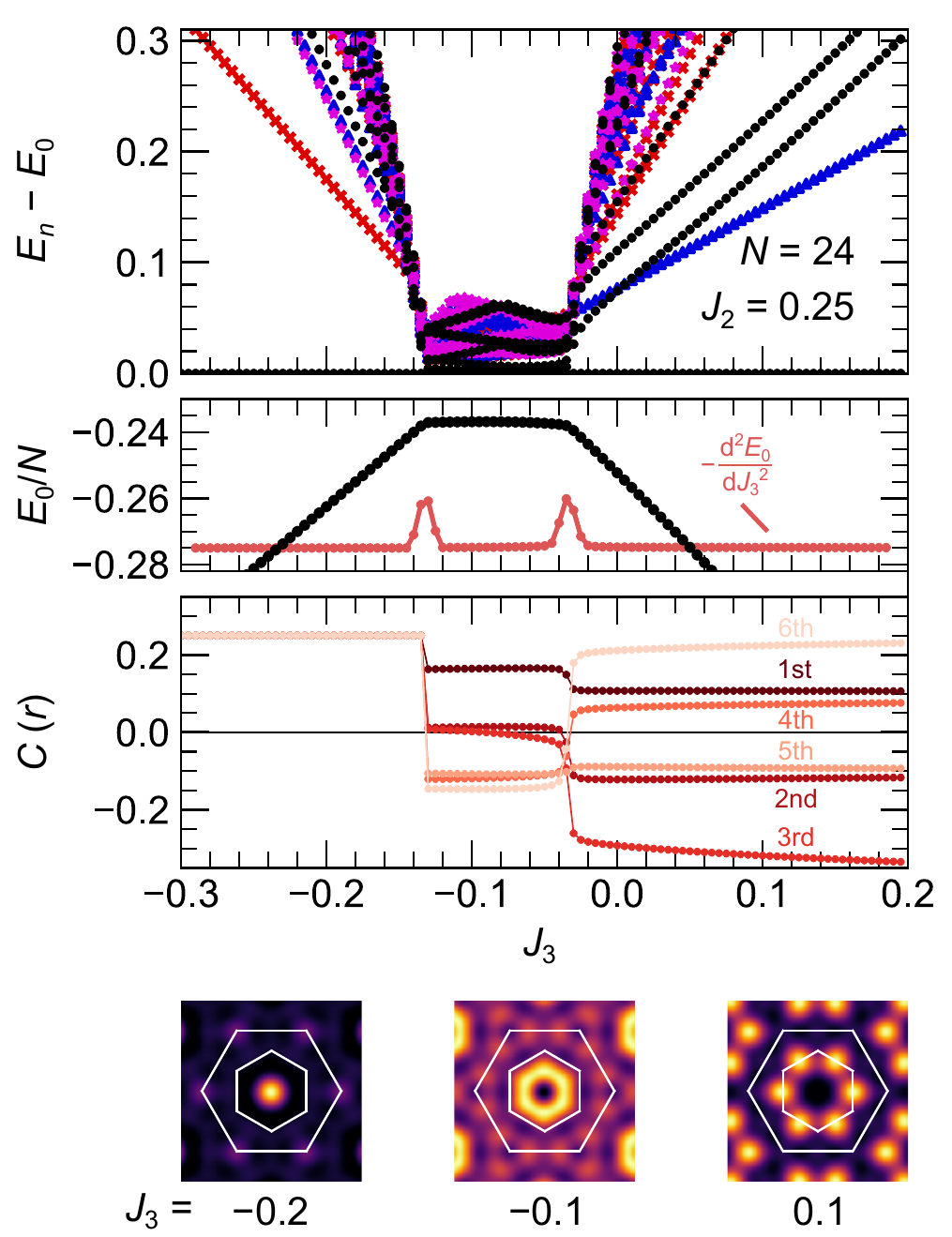}
\caption{
{\bf Stability of the QSL phase for a vertical $J_3$ cut} for fixed $J_2 = 0.25$.
The upper panels show the energy-spectrum, ground state energy (and its second derivative) 
as well as spin-spin correlation functions for varying distances (indicated by the color shading). 
In addition, we show the static structure $S(\bm{q})$ for three values of $J_3$, corresponding
to the ferromagnetic ($J_3=-0.2$), spin liquid ($J_3=-0.1$), and zig-zag phases ($J_3=0.1$).
Note that in contrast to the classical and pf-FRG calculations the ring-like feature in the structure 
factor survives in the spin liquid regime for these ED calculations.
All data were calculated by ED for the $N = 24$ cluster with periodic boundary conditions.
}
\label{fig:J2_cut}
\end{figure}

A vertical cut (varying $J_3$ for fixed $J_2$) is shown in Fig.~\ref{fig:J2_cut}, with again similar signatures in the 
spectrum and ground-state energy. Of particular interest here, might be the spin-spin correlations. In real space, they exhibit
a strong suppression on the level of second and third neighbor spins (where they almost exactly vanish in the
spin liquid regime along this cut), while further neighbor spin-spin correlations are non-negligible. When we look at the spin structure factor in momentum space, we see a ring-like
feature for small momenta -- a feature which is, of course, reminiscent of the spin spiral ring of the classical model.
We note that this feature is not stable for any finite $J_3$ in the classical model nor in our pf-FRG calculations 
pointing to strong quantum fluctuations as the source of this feature.

\section{Energy level spectroscopy}

\begin{figure}[h!]
\includegraphics[width=.9\columnwidth]{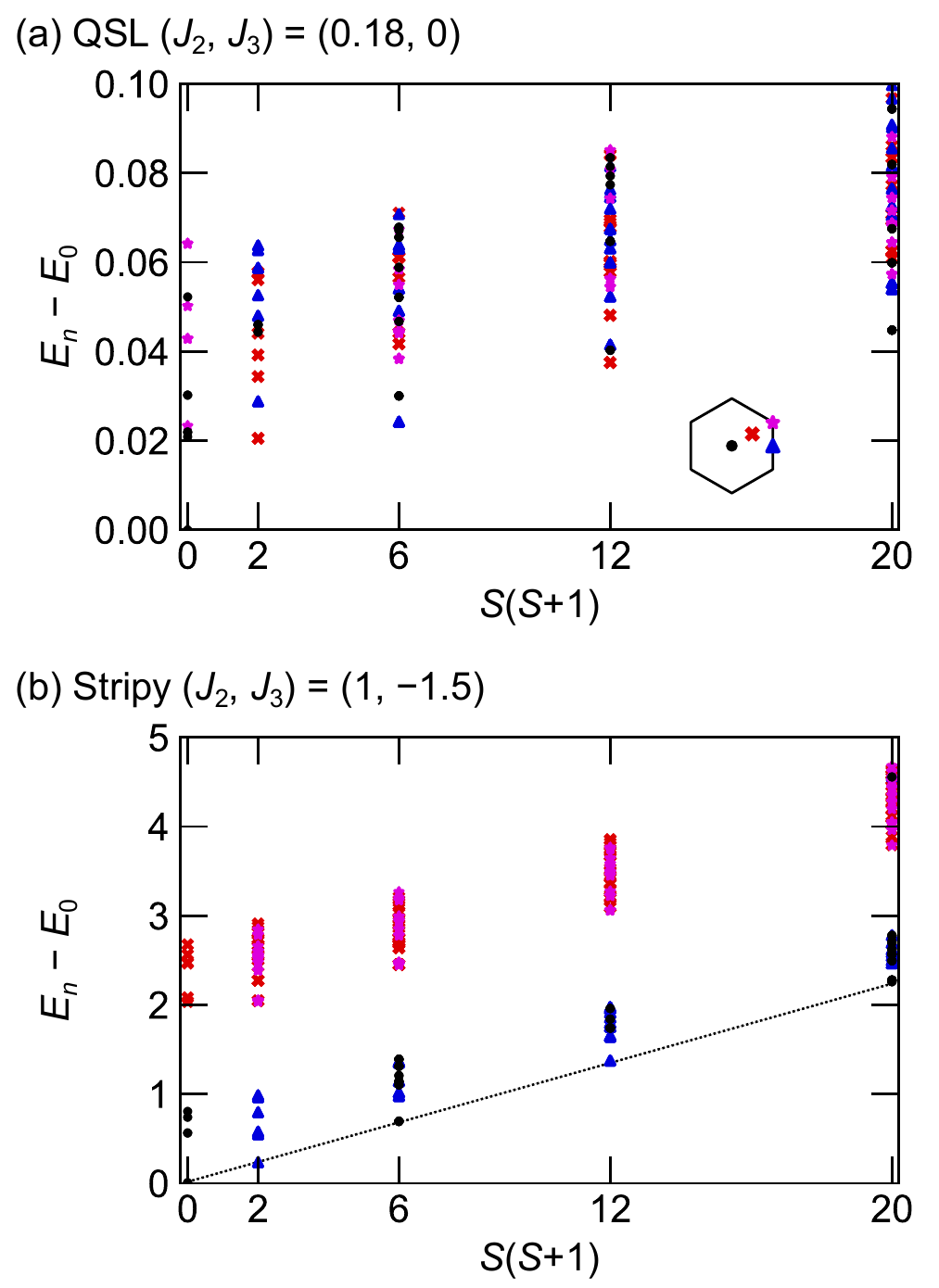}
\caption{
{\bf Energy level spectroscopy} for ED data of $N=24$ system.
The stripy phase (lower panel) shows a clear tower of states \cite{Wietek2017}  indicated by the black line,
while such a structure is absent in the spin liquid phase (upper panel). Note also the significant difference in the energy scales on the vertical axes between the two phases.
}
\label{fig:tos}
\end{figure}

The quantum disordered nature of the spin liquid phase can also be revealed by looking at energy level spectroscopy \cite{Wietek2017}.
In Fig.~\ref{fig:tos} we plot  the energy spectrum, obtained in our ED calculations, against the spin quantum number $S\cdot(S+1)$ 
of the respective eigenstates. The lower panel shows data for the magnetically ordered, stripy phase. Here one can clearly discern
a {\em tower of states} \cite{Wietek2017}, as indicated by the black solid line -- a hierarchy of magnetic excitations expected for any
magnetically ordered state. In contrast, the upper panel shows the energy level spectroscopy for the spin liquid regime. No tower
of states is discernible in this data, and the overall energy scale is suppressed by two orders of magnitude in comparison the magnetically
ordered state in the lower panel (while the coupling strengths differ only by an order of magnitude in the two panels).

\section{Dimer-dimer correlations}

\begin{figure}[t]
\includegraphics[width=.9\columnwidth]{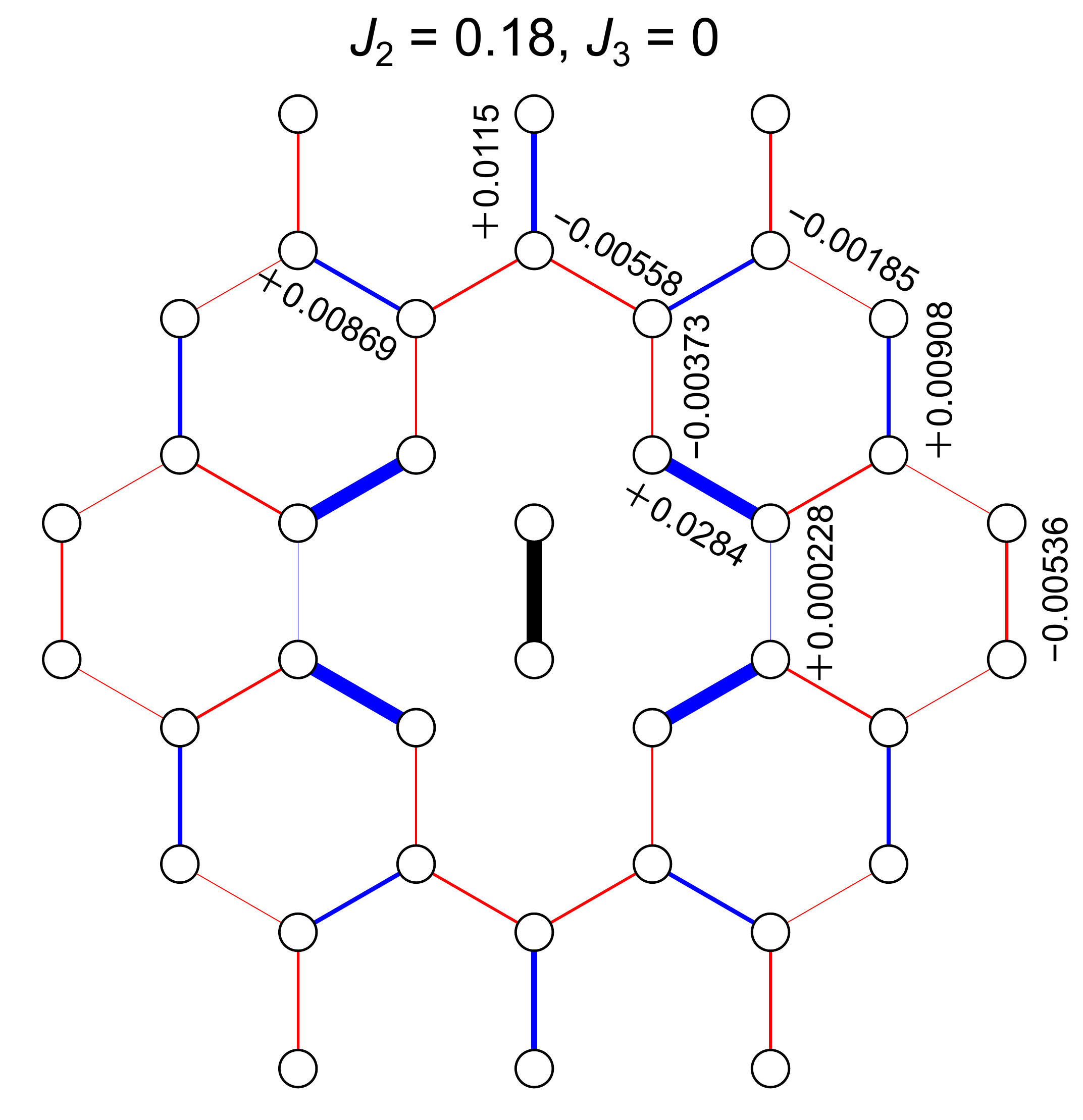}
\caption{
{\bf Dimer-dimer correlations in the QSL phase.} 
Shown are results from ED calculations for the $N = 24$ site cluster with periodic boundary conditions.
}
\label{fig:supp3}
\end{figure}

In the main text, we discuss that one key distinction between anti- and ferromagnetic nearest neighbor interactions is the tendency
of such couplings to favor/disfavor the formation of local singlet dimers. While for the antiferromagnetic case one finds a quantum disordered
regime in which a plaquette valence bond crystal forms (indicated by the purple region in the insets of Fig.~\ref{fig:PD_summary},
no such phase exists for ferromagnetic nearest neighbor couplings. 
This can easily be seen by the absence of any appreciable dimer-dimer correlations as shown in Fig.~\ref{fig:supp3}.

\end{document}